\documentclass[11pt,a4paper]{article}
\usepackage[margin=2.5cm]{geometry}
\usepackage{amsmath}
\usepackage{amssymb}
\usepackage{braket}
\usepackage{graphicx}
\usepackage[colorlinks=true,allcolors=blue]{hyperref}

\title{Resonance-controlled post-selected mode entanglement in one-dimensional quantum scattering}
\author{V\'ictor M Villalba\\[4pt]
{\small School of Mathematics, Monash University, Clayton, VIC 3800, Australia}\\
{\small \href{mailto:victor.villalba2@monash.edu}{victor.villalba2@monash.edu}}}
\date{}

\begin{document}

\maketitle

\begin{abstract}
We study resonance-controlled post-selected mode entanglement in a
one-dimensional quantum-scattering model. Two incoming electrons scatter
independently from a symmetric double-barrier structure and the sector with
one reflected and one transmitted electron is selected. The idealised
delta-function barriers used in earlier work are replaced by exponentially
localised cusp barriers, $U(x)=U_0\exp(-|x\mp d/2|/a)$, for which the single-barrier scattering problem is exactly solvable in terms
of modified Bessel functions of imaginary order. We derive the corresponding
reflection and transmission amplitudes, verify their current normalisation and
validate them independently by direct numerical integration. Two separated
cusp barriers are then combined through a Fabry--P\'erot composition, whose
accuracy is quantified by solving the full two-centre Schr\"odinger equation.
In the well-separated regime the approximation reproduces the full resonance
positions to high accuracy. The concurrence zeros are controlled by the
resonance phase condition $\delta(\kappa)+\kappa d=n\pi$, with a continuously
unwrapped single-barrier reflection phase $\delta$. We also calculate the
post-selection probability $P_{RT}$, showing explicitly that large conditional
concurrence can occur in a low-probability sector. Systematic scans in barrier
width, integrated strength and reference momentum show that finite width can
shift off-resonant concurrence maxima and, depending on the reference
momentum, reduce their attainable height. The results identify
resonance-controlled conditional entanglement as a robust feature of the
well-separated double-barrier geometry rather than an artefact of the
zero-range idealisation.
\end{abstract}

\medskip
\noindent\textbf{Keywords:} quantum entanglement, concurrence, entanglement entropy, one-dimensional scattering, resonant tunnelling, cusp potential, quantum transport, quantum information

\section{Introduction}

The production of entanglement between two otherwise independent
electrons via elastic scattering in a mesoscopic conductor has been
studied extensively as a route to on-demand, flying-qubit
entanglement sources~\cite{Martin2003,Beenakker2003,Schliemann2001},
a research direction that continues to develop actively: recent
reviews of the semiconductor flying-qubit platform document steady
experimental progress~\cite{Edlbauer2022}, and controlled two-electron
collisions at a mesoscopic beam splitter have now been realised
experimentally~\cite{Ubbelohde2023}, with closely related theoretical
work analysing how Coulomb interaction between the two electrons
modifies barrier partitioning~\cite{Ryu2022}. Scattering-induced entanglement has also been studied in complementary
settings, including scattering of particles from localised spin
impurities~\cite{Ghanbari2013Spin}, finite-width Gaussian wave packets
incident on localised impurities~\cite{Ghanbari2014Wavepackets}, direct
collisions of initially uncorrelated particles~\cite{Ghanbari2014Collisions},
electronic waveguides containing static magnetic impurities~\cite{Ghanbari2016Waveguides},
and Klein tunnelling in graphene~\cite{Ghanbari2016Graphene}. These studies
provide useful points of comparison because they likewise connect the
entanglement of outgoing alternatives to scattering amplitudes, while also
highlighting the importance of wave-packet width, spin, mode structure and
the choice of subsystem partition. Previous work~\cite{Lopez2006} showed that a
quasi-one-dimensional, two-channel conductor containing a symmetric
double-barrier scattering region produces a concurrence $\eta$ between
two incoming electrons of momenta $\kappa_1,\kappa_2$ that vanishes
sharply whenever either electron is tuned through a transmission
resonance of the double-barrier structure, and is enhanced between
resonances. That analysis, like most exactly solvable treatments of
resonant tunnelling, idealised each barrier as a zero-range
delta-function potential, $V(x)=\lambda\,\delta(x\mp d/2)$.

The delta-function idealisation is convenient but not realistic: a
gate-defined or lithographically patterned tunnel barrier has a finite
spatial extent set by the confinement length scale of the device. It
is natural to ask whether the sharp vanishing of the conditional concurrence at resonance
survives when the barrier is given a finite width, and whether the
resonance positions and the concurrence between zeros are sensitive to
the detailed shape of the barrier.

The present work is addressed to this question from the point of view of
one-dimensional quantum scattering and quantum-information generation.
The central observable is the concurrence of the post-selected outgoing
state in which one electron is reflected and the other transmitted.
Thus the role of the barrier is not only to provide a solvable scattering
problem, but also to control the relative weights of the two alternatives
that form the entangled two-mode state. Transmission resonances are then
expected to suppress the concurrence, while off-resonant scattering can
produce nearly balanced reflected/transmitted amplitudes and hence large
entanglement.

The purpose of replacing the delta barriers by finite-width cusp barriers
is to test the structural stability of this resonance-controlled
entanglement mechanism. If the concurrence zeros and the peak structure
were artefacts of the zero-range idealisation, they would be strongly
shifted or washed out when the barrier is given a finite spatial extent.
Conversely, if the same qualitative behaviour persists for an exactly
solvable finite-width barrier, the mechanism can be regarded as a robust
feature of resonant one-dimensional scattering rather than a peculiarity
of the delta model.

Here we address this question using the cusp-shaped potential
\begin{equation}
U(x)=U_0\,e^{-|x|/a},
\end{equation}
centred at each barrier location. The cusp potential is continuous everywhere but has a discontinuous
derivative (a ``kink'') at its peak, whereas the delta potential is a
singular distribution concentrated at a point rather than an ordinary
function. It reduces to a delta-function barrier of strength
$\lambda=2U_0a$ in the limit $a\to0$ at fixed $\lambda$, and for finite
$a$ it is exactly solvable: the Schr\"odinger equation on each half-line
reduces, via the substitution
$z=2a\sqrt{U_0}\,e^{-|x|/(2a)}$, to the modified Bessel equation of
imaginary order $\nu=2ia\kappa$. This makes the cusp barrier one of the few exponentially localised
barriers for which the single-barrier scattering problem is available in
closed form. For two cusp barriers the analytical Fabry--P\'erot composition
is used only in the well-separated regime and is validated below against
direct numerical integration of the full two-centre problem.

\section{Single-barrier scattering amplitudes}

\subsection{Reduction to the modified Bessel equation}

Consider a particle of wavenumber $\kappa$ (we set $\hbar=2m=1$
throughout) incident on a single cusp barrier centred at the origin,
$U(x)=U_0e^{-|x|/a}$, $U_0>0$. By symmetry it suffices to work on the
half-line $x>0$ and use $x\to-x$ afterwards for $x<0$. Writing
$s=|x|\ge0$, the stationary Schr\"odinger equation is
\begin{equation}
-\psi''(s)+U_0e^{-s/a}\psi(s)=\kappa^2\psi(s).
\label{eq:sch}
\end{equation}
Introduce the new independent variable
\begin{equation}
z(s)=\mu\,e^{-s/(2a)},\qquad \mu\equiv2a\sqrt{U_0},
\end{equation}
so that $z\to\mu$ at $s=0$ and $z\to0$ as $s\to\infty$, and write
$\psi(s)=f(z(s))$. Using $dz/ds=-z/(2a)$,
\begin{equation}
\frac{d\psi}{ds}=-\frac{z}{2a}f'(z),\qquad
\frac{d^2\psi}{ds^2}=\frac{z^2}{4a^2}f''(z)+\frac{z}{4a^2}f'(z),
\end{equation}
and noting $U_0e^{-s/a}=U_0(z/\mu)^2=z^2/(4a^2)$ (using
$\mu^2=4a^2U_0$), Eq.~(\ref{eq:sch}) becomes, after multiplying
through by $-4a^2$,
\begin{equation}
z^2f''(z)+zf'(z)-\bigl(z^2-4a^2\kappa^2\bigr)f(z)=0.
\label{eq:modbessel}
\end{equation}
This is the \emph{modified} Bessel equation,
$z^2f''+zf'-(z^2+\nu^2)f=0$, of order
\begin{equation}
\nu=2ia\kappa,
\end{equation}
since $\nu^2=-4a^2\kappa^2$. Its two linearly independent solutions
are the modified Bessel functions $I_{\nu}(z)$ and $I_{-\nu}(z)$
(these are independent whenever $\nu$ is not an integer, which holds
here since $\nu$ is purely imaginary and nonzero for $\kappa\ne0$). Note
that the ordinary Bessel functions $J_{\pm\nu}(z)$, which solve
$z^2f''+zf'+(z^2-\nu^2)f=0$, do not solve Eq.~(\ref{eq:sch}), although
they share the same leading small-$z$ behaviour.

\subsection{Asymptotic identification of incident, reflected and transmitted waves}

The behaviour of $I_{\pm\nu}(z)$ as $z\to0$ is
\begin{equation}
I_{\pm\nu}(z)\;\longrightarrow\;\frac{1}{\Gamma(1\pm\nu)}
\left(\frac{z}{2}\right)^{\pm\nu},
\end{equation}
the same leading power-law behaviour as the ordinary Bessel function
$J_{\pm\nu}(z)$ (the two differ only in subleading terms in $z^2$).
Substituting $z=z(s)=\mu e^{-s/(2a)}$ and $\nu=2ia\kappa$,
\begin{equation}
\left(\frac{z(s)}{2}\right)^{\nu}
=\exp\!\Bigl[2ia\kappa\ln\bigl(\tfrac{\mu}{2}\bigr)-i\kappa s\Bigr]
\;\propto\;e^{-i\kappa s},
\end{equation}
so that, up to an $s$-independent constant, $I_\nu(z(s))\to e^{-i\kappa
s}$ as $s\to\infty$, while $I_{-\nu}(z(s))\to e^{+i\kappa s}$. In terms
of $x$: on the left half-line ($x<0$, $s=-x$) the combination
$e^{-i\kappa s}=e^{i\kappa x}$ is a wave travelling in the $+x$
direction (the physical incident wave from $x=-\infty$), and
$e^{+i\kappa s}=e^{-i\kappa x}$ travels in the $-x$ direction (the
reflected wave); on the right half-line ($x>0$, $s=x$),
$e^{-i\kappa s}=e^{-i\kappa x}$ would represent an unphysical wave
incoming from $x=+\infty$, which must be absent, while
$e^{+i\kappa s}=e^{i\kappa x}$ is the physical outgoing transmitted
wave.

We therefore write, with
\[
C_{\pm\nu}\equiv\frac{(\mu/2)^{\pm\nu}}{\Gamma(1\pm\nu)},
\]
\begin{align}
\psi_L(x) &= \frac{1}{C_\nu}\,I_\nu(z(-x))
  +\frac{r_0}{C_{-\nu}}\,I_{-\nu}(z(-x)), && x<0,
  \label{eq:psiL}\\
\psi_R(x) &= \frac{t_0}{C_{-\nu}}\,I_{-\nu}(z(x)), && x>0.
  \label{eq:psiR}
\end{align}
By construction,
\[
\psi_L(x)\longrightarrow e^{i\kappa x}+r_0e^{-i\kappa x}
\quad (x\to-\infty),\qquad
\psi_R(x)\longrightarrow t_0e^{i\kappa x}
\quad (x\to+\infty).
\]
Thus Eqs.~(\ref{eq:psiL})--(\ref{eq:psiR}) use unit incident
\emph{amplitude}.  The corresponding flux normalisation is also
transparent.  In the units $\hbar=2m=1$, the probability current is
\begin{equation}
j=-i\left(\psi^*\psi'-\psi\psi'^*\right)
  =2\,\mathrm{Im}\!\left(\psi^*\psi'\right).
\label{eq:current}
\end{equation}
A plane wave $A e^{\pm i\kappa x}$ therefore carries current
$\pm 2\kappa |A|^2$.  Hence
\begin{equation}
j_{\rm in}=2\kappa,\qquad
j_{\rm ref}=-2\kappa|r_0|^2,\qquad
j_{\rm tr}=2\kappa|t_0|^2.
\label{eq:asymptoticcurrents}
\end{equation}
If a unit-flux rather than a unit-amplitude normalisation is desired,
the entire scattering state may simply be multiplied by
$(2\kappa)^{-1/2}$; the amplitudes $r_0$ and $t_0$ are unchanged.
Because the asymptotic wavenumber is the same on both sides of the
barrier, the reflection and transmission probabilities are therefore
$R_0=|r_0|^2$ and $T_0=|t_0|^2$.

\subsection{Matching at the cusp and the closed-form amplitudes}

The potential $U(x)$ is continuous at $x=0$, so both $\psi$ and
$d\psi/dx$ are continuous.  The signs in the derivative matching are
most easily seen by writing the Bessel arguments explicitly on the two
half-lines:
\[
z_L(x)=\mu e^{x/(2a)},\qquad x<0,
\qquad
z_R(x)=\mu e^{-x/(2a)},\qquad x>0.
\]
Consequently
\[
\frac{dz_L}{dx}=+\frac{z_L}{2a},\qquad
\frac{dz_R}{dx}=-\frac{z_R}{2a}.
\]
At the cusp, $z_L(0)=z_R(0)=\mu$, and therefore
\begin{align}
\psi_L'(0^-)
 &=\frac{\mu}{2a}\left[
 \frac{I_\nu'(\mu)}{C_\nu}
 +\frac{r_0 I_{-\nu}'(\mu)}{C_{-\nu}}\right],\\
\psi_R'(0^+)
 &=-\frac{\mu}{2a}\,
 \frac{t_0 I_{-\nu}'(\mu)}{C_{-\nu}}.
\end{align}
Here and below a prime on $I_{\pm\nu}$ denotes differentiation with
respect to the Bessel-function argument, whereas a prime on $\psi$
denotes differentiation with respect to $x$.

The continuity conditions $\psi_L(0)=\psi_R(0)$ and
$\psi_L'(0^-)=\psi_R'(0^+)$ then give
\begin{align}
\frac{I_\nu(\mu)}{C_\nu}
+\frac{r_0I_{-\nu}(\mu)}{C_{-\nu}}
&=
\frac{t_0I_{-\nu}(\mu)}{C_{-\nu}},
\label{eq:match1}\\
\frac{I_\nu'(\mu)}{C_\nu}
+\frac{r_0I_{-\nu}'(\mu)}{C_{-\nu}}
&=
-\frac{t_0I_{-\nu}'(\mu)}{C_{-\nu}}.
\label{eq:match2}
\end{align}
It is convenient to define
\begin{equation}
K\equiv\frac{C_{-\nu}}{C_\nu}
=
\frac{\Gamma(1+\nu)}{\Gamma(1-\nu)}
\left(\frac{\mu}{2}\right)^{-2\nu}.
\label{eq:Kcorrect}
\end{equation}
Multiplying Eqs.~(\ref{eq:match1})--(\ref{eq:match2}) by
$C_{-\nu}$ gives directly
\begin{equation}
r_0-t_0=-K\,\frac{I_\nu(\mu)}{I_{-\nu}(\mu)},\qquad
r_0+t_0=-K\,\frac{I_\nu'(\mu)}{I_{-\nu}'(\mu)}.
\label{eq:rplustminus}
\end{equation}
Solving these two equations yields
\begin{equation}
\boxed{
r_0(\kappa)=
-\frac{K}{2}\left[
\frac{I_\nu(\mu)}{I_{-\nu}(\mu)}
+
\frac{I_\nu'(\mu)}{I_{-\nu}'(\mu)}
\right],
\qquad
t_0(\kappa)=
\frac{K}{2}\left[
\frac{I_\nu(\mu)}{I_{-\nu}(\mu)}
-
\frac{I_\nu'(\mu)}{I_{-\nu}'(\mu)}
\right].
}
\label{eq:r0t0}
\end{equation}
Note that the ratio entering $K$ is $C_{-\nu}/C_\nu$, which follows
directly from multiplying the matching conditions by $C_{-\nu}$.

The flux relation can also be verified analytically from the Bessel
representation.  For real $\mu$ and purely imaginary
$\nu=2ia\kappa$,
\[
C_{-\nu}=C_\nu^*,\qquad
I_{-\nu}(\mu)=I_\nu(\mu)^*,\qquad
I_{-\nu}'(\mu)=I_\nu'(\mu)^*.
\]
Hence each of
\[
K,\qquad
A_\nu\equiv\frac{I_\nu(\mu)}{I_{-\nu}(\mu)},\qquad
B_\nu\equiv\frac{I_\nu'(\mu)}{I_{-\nu}'(\mu)}
\]
has unit modulus.  Since
\[
r_0=-\frac{K}{2}(A_\nu+B_\nu),\qquad
t_0=\frac{K}{2}(A_\nu-B_\nu),
\]
we obtain
\begin{align}
|r_0|^2+|t_0|^2
&=\frac14\left(|A_\nu+B_\nu|^2+
|A_\nu-B_\nu|^2\right)\\
&=\frac12\left(|A_\nu|^2+|B_\nu|^2\right)=1.
\label{eq:unitarity_analytic}
\end{align}
Together with Eq.~(\ref{eq:asymptoticcurrents}), this is precisely
current conservation, $j_{\rm in}+j_{\rm ref}=j_{\rm tr}$.

\subsection{Independent numerical validation of the single-cusp amplitudes}

As an independent check that does not use Bessel functions, we integrated
the Schr\"odinger equation with $U(x)=U_0e^{-|x|/a}$ numerically with an adaptive high-order Runge--Kutta
scheme over $|x|\le30a$, where the residual potential is below
$10^{-13}U_0$, and matched to plane waves at the endpoints.

Table~\ref{tab:single-cusp-validation} compares the complex amplitudes
obtained in this completely independent numerical calculation with
Eq.~(\ref{eq:r0t0}).  The quoted errors are maxima over 37 equally
spaced values of $\kappa$ in $0.2\leq\kappa\leq2.0$, at
$\lambda=2U_0a=0.01$.  The agreement is at the $10^{-11}$ level in the
complex amplitudes, while the numerical flux residual
$||r_0|^2+|t_0|^2-1|$ remains below $4\times10^{-13}$.

\begin{table}[htbp]
\centering
\caption{Independent numerical validation of the single-cusp
reflection and transmission amplitudes.  The numerical calculation
integrates the original Schr\"odinger equation directly and does not
use the Bessel-function solution.}
\label{tab:single-cusp-validation}
\begin{tabular}{cccc}
\hline
$a$ &
$\max|r_0^{\rm ana}-r_0^{\rm num}|$ &
$\max|t_0^{\rm ana}-t_0^{\rm num}|$ &
$\max\bigl||r_0^{\rm num}|^2+|t_0^{\rm num}|^2-1\bigr|$\\
\hline
0.3 & $8.6\times10^{-12}$ & $9.3\times10^{-12}$ & $1.6\times10^{-13}$\\
0.6 & $1.7\times10^{-11}$ & $1.7\times10^{-11}$ & $3.1\times10^{-13}$\\
\hline
\end{tabular}
\end{table}

\subsection{Delta-barrier limit and convergence rate}
\label{sec:delta-limit}

The comparison with the delta barrier is made at fixed integrated
strength
\begin{equation}
\lambda=\int_{-\infty}^{\infty}U_a(x)\,dx,
\qquad
U_a(x)=\frac{\lambda}{2a}e^{-|x|/a}.
\label{eq:Ua_delta_family}
\end{equation}
This prescription is natural because $U_a$ is an even approximate
identity of total weight $\lambda$.  Indeed, for a smooth test function
$\varphi$,
\begin{align}
\int_{-\infty}^{\infty}U_a(x)\varphi(x)\,dx
&=
\frac{\lambda}{2}\int_{-\infty}^{\infty}
e^{-|y|}\varphi(ay)\,dy \nonumber\\
&=
\lambda\left[
\varphi(0)+a^2\varphi''(0)+O(a^4)
\right],
\label{eq:distributional_delta_limit}
\end{align}
where the odd terms vanish by symmetry and
$\int e^{-|y|}dy=2$,
$\int y^2e^{-|y|}dy=4$.  Thus
\begin{equation}
U_a
=
\lambda\delta
+
\lambda a^2\delta''
+
O_{\mathcal D'}(a^4)
\label{eq:distribution_expansion}
\end{equation}
in the distributional sense, for sufficiently smooth test functions.
Equation~(\ref{eq:distributional_delta_limit}) explains both why
matching the integrated strength is the natural zero-range comparison
and why the potential itself approaches $\lambda\delta(x)$
distributionally. Other prescriptions, such as matching the single-barrier
reflection probability or phase at a chosen energy, would define different
finite-$a$ comparisons; the present choice is the one tied to the
distributional delta limit and is independent of energy.

The distributional $O(a^2)$ correction in
Eq.~(\ref{eq:distribution_expansion}) does \emph{not}, however, imply
$O(a^2)$ convergence of the scattering amplitudes.  In the delta
limit the scattering wavefunction is continuous but its derivative
develops the finite jump
$\psi'(0^+)-\psi'(0^-)=\lambda\psi(0)$, so the limiting scattering
state is not a smooth test function at the origin.  The convergence
rate of $r_0$ and $t_0$ must therefore be obtained from the scattering
solution itself.

This rate can be derived directly from the exact Bessel representation.
At fixed $\lambda$,
\begin{equation}
q\equiv\frac{\mu^2}{4}=\frac{\lambda a}{2},
\qquad
\nu=2ia\kappa ,
\end{equation}
so both $q$ and $\nu$ are $O(a)$.  Writing
\begin{equation}
I_\nu(\mu)
=
C_\nu S_\nu(q),\qquad
S_\nu(q)
=
1+\frac{q}{1+\nu}
+\frac{q^2}{2(1+\nu)(2+\nu)}
+O(a^3),
\label{eq:Snu_expansion}
\end{equation}
with
$C_\nu=(\mu/2)^\nu/\Gamma(1+\nu)$, the combinations entering
Eq.~(\ref{eq:r0t0}) simplify because
$K=C_{-\nu}/C_\nu$.  In particular,
\begin{equation}
K\frac{I_\nu}{I_{-\nu}}
=
\frac{S_\nu}{S_{-\nu}}
=
1+O(a^2).
\label{eq:A_small_a}
\end{equation}
For the derivative ratio, define
\begin{equation}
D_\nu
\equiv
\nu S_\nu+2q\frac{\partial S_\nu}{\partial q}.
\end{equation}
Since
$I_\nu'(\mu)=C_\nu D_\nu/\mu$, one has
\begin{equation}
K\frac{I_\nu'}{I_{-\nu}'}
=
\frac{D_\nu}{D_{-\nu}}.
\end{equation}
Using Eq.~(\ref{eq:Snu_expansion}),
\begin{align}
D_\nu
&=
a(\lambda+2i\kappa)
+a^2\left(\frac{\lambda^2}{4}-i\lambda\kappa\right)
+O(a^3),\\
D_{-\nu}
&=
a(\lambda-2i\kappa)
+a^2\left(\frac{\lambda^2}{4}+i\lambda\kappa\right)
+O(a^3).
\end{align}
Substitution into Eq.~(\ref{eq:r0t0}) gives the explicit small-$a$
expansion
\begin{align}
r_0^{\rm cusp}(\kappa)
&=
\frac{\lambda}{2i\kappa-\lambda}
+
\frac{3i\kappa\lambda^2}
     {2(2i\kappa-\lambda)^2}\,a
+O(a^2),
\label{eq:r_small_a}\\
t_0^{\rm cusp}(\kappa)
&=
\frac{2i\kappa}{2i\kappa-\lambda}
+
\frac{3i\kappa\lambda^2}
     {2(2i\kappa-\lambda)^2}\,a
+O(a^2).
\label{eq:t_small_a}
\end{align}
The leading terms are precisely the delta-barrier amplitudes.  The
first finite-width correction is linear in $a$, not quadratic.  For
weak barriers this linear term is proportional to $\lambda^2$, which
explains why an apparent faster convergence can occur over a moderate
range of $a$ before the true asymptotic regime is reached.

\begin{figure}[htbp]
\centering
\includegraphics[width=0.76\textwidth]{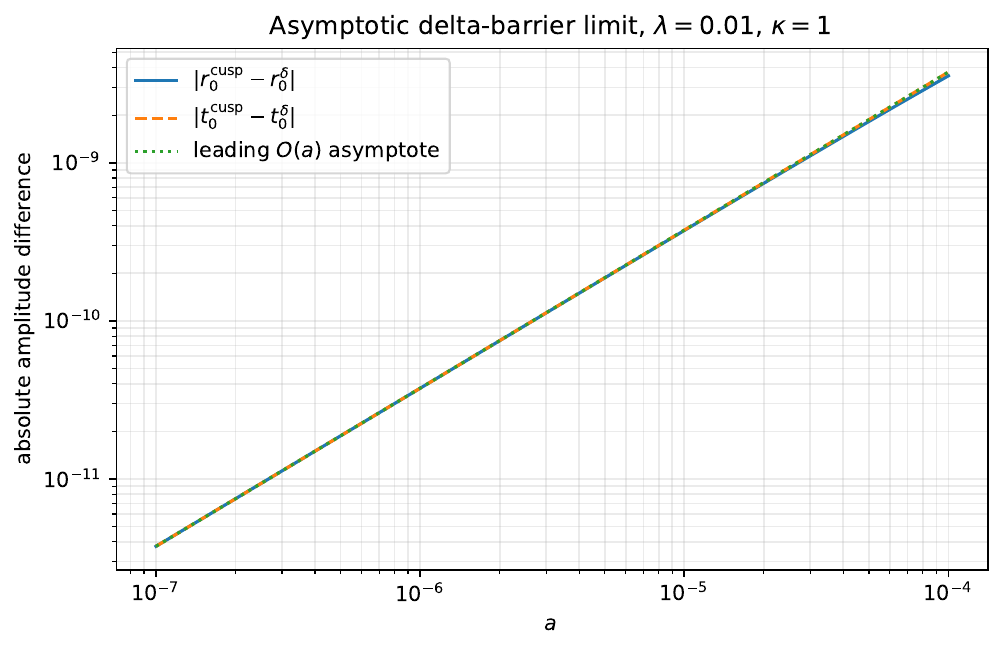}
\caption{Approach of the exact single-cusp amplitudes to the
delta-barrier amplitudes at fixed integrated strength
$\lambda=0.01$ and $\kappa=1$.  The dotted curve is the magnitude of
the leading linear correction in
Eqs.~(\ref{eq:r_small_a})--(\ref{eq:t_small_a}).  Both complex
amplitude differences become asymptotically proportional to $a$.}
\label{fig:delta-limit-convergence}
\end{figure}

Figure~\ref{fig:delta-limit-convergence} confirms
Eqs.~(\ref{eq:r_small_a})--(\ref{eq:t_small_a}) numerically: at fixed
$\lambda=0.01$, log--log fits of $|r_0^{\rm cusp}-r_0^{\delta}|$ and
$|t_0^{\rm cusp}-t_0^{\delta}|$ over $10^{-7}\le a\le10^{-5}$ give exponents
between $0.998$ and $1.000$ for $0.25\le\kappa\le1.5$, as expected for an
$O(a)$ correction.

The expansion in Eqs.~(\ref{eq:r_small_a})--(\ref{eq:t_small_a}) is
asymptotic in $a$ at fixed $\lambda$, and its $O(a)$ coefficient is
proportional to $\lambda^2$. For the barrier widths used in the results,
$\kappa a=O(1)$, a complementary expansion in the weak-barrier limit is
more informative. To first order in $\lambda$ (Born approximation) the
reflection amplitude is proportional to the Fourier transform of the
potential at momentum transfer $2\kappa$,
\[
r_0^{\rm cusp}(\kappa)\simeq-\frac{i}{2\kappa}\int_{-\infty}^{\infty}
U_a(x)\,e^{2i\kappa x}\,dx
=\frac{-i\lambda}{2\kappa}\,\frac{1}{1+4\kappa^2a^2},
\]
valid for all $\kappa a$ when $\lambda\ll\kappa$. The leading delta-barrier
result, $r_0^{\delta}\simeq-i\lambda/(2\kappa)$, is recovered for $a\to0$.
This form-factor result shows that, at weak coupling, finite width
reduces the reflection \emph{modulus} by the form factor
$(1+4\kappa^2a^2)^{-1}$ while leaving the reflection \emph{phase} unchanged
at first order in $\lambda$. For $\lambda=0.01$ and $a=0.3$ the exact
amplitudes of Eq.~(\ref{eq:r0t0}) give $|r_0^{\rm cusp}|/|r_0^{\delta}|=0.737$
at $\kappa=1$ and $0.564$ at $\kappa=1.47$, compared with the form-factor
values $0.735$ and $0.562$, whereas the reflection phases at these momenta differ from the
delta values by less than $10^{-5}$~rad ($6.8\times10^{-6}$ and
$3.1\times10^{-6}$~rad, respectively). This explains quantitatively the
two features found below: because the resonance condition,
Eq.~(\ref{eq:resonance}), depends only on the phase $\delta(\kappa)$, the
cusp and delta resonance positions differ only at higher order in $\lambda$
(Table~\ref{tab:lambda-resonance-shifts}); because the concurrence depends
on the ratios of reflection moduli at $\kappa_1$ and $\kappa_2$, the
$\kappa$-dependent form factor modifies the off-resonant concurrence maxima
(Fig.~\ref{fig:concurrence} and Table~\ref{tab:a-dependence}).

\section{Two-barrier composition and the resonance condition}

\subsection{The Fabry--P\'erot composition}

For two identical barriers centred at $x=\mp d/2$, the composite
amplitudes can be built from the single-barrier amplitudes $r_0,t_0$
of Sec.~2 (identical for both barriers by translation symmetry) by
summing the geometric series of multiple reflections in the region
between them. A wave transmitted through the left barrier with
amplitude $t_0$ propagates a distance $d$, acquiring a phase
$e^{i\kappa d}$, before reaching the right barrier; there it is either
transmitted out (amplitude $t_0$, contributing the leading term
$t_0\,e^{i\kappa d}\,t_0$ to the total transmission) or reflected back
(amplitude $r_0$) toward the left barrier, where it can reflect again
(amplitude $r_0$) and repeat the traversal, contributing an extra
factor $r_0^2e^{2i\kappa d}$ per round trip. Summing the resulting
geometric series,
\begin{equation}
t(\kappa)=t_0e^{i\kappa d}t_0\sum_{n=0}^{\infty}
\bigl(r_0^2e^{2i\kappa d}\bigr)^n
=\frac{t_0^2\,e^{i\kappa d}}{1-r_0^2\,e^{2i\kappa d}},
\end{equation}
and similarly, tracking the paths that return to the left of the
structure (direct reflection off the left barrier, plus paths that
cross into the well, reflect off the right barrier, and re-emerge
through the left barrier after any number of additional round trips),
\begin{equation}
r(\kappa)=r_0+t_0\,e^{i\kappa d}\,r_0\,e^{i\kappa d}\,t_0
\sum_{n=0}^{\infty}\bigl(r_0^2e^{2i\kappa d}\bigr)^n
=r_0+\frac{t_0^2\,r_0\,e^{2i\kappa d}}{1-r_0^2\,e^{2i\kappa d}}.
\end{equation}
This Fabry--P\'erot (Airy) composition is the standard tool for
combining well-separated one-dimensional scatterers into a single
effective $S$-matrix; standard treatments of the method and of
resonant tunnelling through composite barrier structures can be found
in Refs.~\cite{Razavy2003,MelloKumar2004}. The composition formula is
barrier-shape-independent in form; in the well-separated-barrier regime
considered here, it applies equally to the delta and cusp cases once
$r_0,t_0$ are specified from Sec.~2. We have confirmed
$|r|^2+|t|^2=1$ for both barrier shapes, as required since the
composite structure is again a real potential.

\subsection{Numerical validation of the well-separated-scatterer approximation}
\label{sec:validity}

The geometric-series argument of the previous subsection treats the two
barriers as independent scatterers connected by free propagation over the
intervening distance $d$. For delta-function barriers this composition is
exact for every $d>0$. For the cusp barrier, however,
$U(x)=U_0e^{-|x|/a}$ has infinite support, so the full two-centre potential
\begin{equation}
U_{\rm 2B}(x)=U_0e^{-|x+d/2|/a}+U_0e^{-|x-d/2|/a}
\label{eq:full-two-centre}
\end{equation}
contains overlapping exponential tails throughout the region between the two
centres. The Fabry--P\'erot construction is therefore an approximation for the
cusp case, expected to become accurate when $d/a$ is large.

To test this directly, we solved the Schr\"odinger equation for the full
potential, Eq.~(\ref{eq:full-two-centre}), by the same numerical integration and
plane-wave matching, now over $|x|\le d/2+30a$. This calculation retains the
overlapping tails and does not assume a field-free region between the
barriers.

The resonance locations were obtained independently in the Fabry--P\'erot and
full two-centre descriptions by minimising $R(\kappa)=|r(\kappa)|^2$. At
fixed $\lambda=0.01$ and $d=2\pi$, Table~\ref{tab:resonance-validation}
shows the largest absolute displacement among the first three resonances as
$a$ is increased. The principal case used for the main concurrence plots,
$a=0.3$ ($d/a\simeq20.94$), gives a maximum shift of only
$1.6\times10^{-9}$ in $\kappa$, whereas the error rises to about
$3.0\times10^{-5}$ when $d/a\simeq2.99$.

\begin{table}[htbp]
\centering
\caption{Numerical validation of the Fabry--P\'erot resonance
positions against direct integration of the full two-centre cusp potential.
Here $\lambda=0.01$ and $d=2\pi$ are fixed while $a$ is varied. The quoted
error is the largest absolute displacement among the first three resonances.}
\label{tab:resonance-validation}
\begin{tabular}{ccc}
\hline
$a$ & $d/a$ & $\max_{n=1,2,3}|\kappa_{n,\rm full}-\kappa_{n,\rm FP}|$ \\
\hline
0.3 & 20.94 & $1.56\times10^{-9}$ \\
0.6 & 10.47 & $4.44\times10^{-8}$ \\
0.9 & 6.98  & $1.31\times10^{-6}$ \\
1.2 & 5.24  & $6.60\times10^{-6}$ \\
1.8 & 3.49  & $2.49\times10^{-5}$ \\
2.1 & 2.99  & $3.01\times10^{-5}$ \\
\hline
\end{tabular}
\end{table}

The approximation was also checked at the level of the resonance width,
transmission probability and conditional concurrence. Because the weak
barriers have a slowly varying background transmission close to unity, we
define the width of the second resonance by a local half-contrast criterion:
on each side of the resonance we first locate the neighbouring local minimum
of $T(\kappa)$ and then determine the crossing halfway between that minimum
and the resonant maximum. The sum of the two half-widths defines $\Gamma_2$.
This avoids imposing an artificial zero-transmission background. For three
representative separations, Table~\ref{tab:representative-validation}
compares $\Gamma_2$ and also reports sampled maximum differences in
transmission and concurrence. The transmission differences were sampled at
12 equally spaced values in $0.18\le\kappa\le1.45$; the concurrence
comparison used $\kappa_1=1$ and 12 equally spaced values of
$0.12\le\Delta\kappa\le1.40$. These sampled errors are practical validation
diagnostics rather than rigorous global suprema.

\begin{table*}[htbp]
\centering
\caption{Representative validation of resonance width, transmission and
conditional concurrence for the full two-centre cusp potential. The relative
width error is $|\Gamma_{2,\rm full}-\Gamma_{2,\rm FP}|/\Gamma_{2,\rm full}$.
The last column reports the sampled transmission and concurrence errors over
the ranges specified in the text.}
\label{tab:representative-validation}
\resizebox{\textwidth}{!}{%
\begin{tabular}{cccccc}
\hline
$a$ & $d/a$ & $\Gamma_{2,\rm FP}$ & $\Gamma_{2,\rm full}$ & relative width error & $\max|\Delta T|\;/\;\max|\Delta\eta|$ \\
\hline
0.3 & 20.94 & 0.27255441 & 0.27255442 & $1.33\times10^{-8}$ & $8.1\times10^{-13}\;/\;2.9\times10^{-9}$ \\
1.2 & 5.24  & 0.30681913 & 0.30681600 & $1.02\times10^{-5}$ & $1.7\times10^{-7}\;/\;1.3\times10^{-5}$ \\
2.1 & 2.99  & 0.33074155 & 0.33073562 & $1.79\times10^{-5}$ & $8.6\times10^{-7}\;/\;1.8\times10^{-4}$ \\
\hline
\end{tabular}}
\end{table*}

The resonance positions, widths, transmission and conditional concurrence
therefore converge consistently to the separated-scatterer result as $d/a$
increases. In the remainder of the paper we use the analytical Fabry--P\'erot
expression to expose the resonance mechanism and the full two-centre
numerical solution when assessing the approximation quantitatively.

\subsection{Resonance condition}

Write the single-barrier reflection amplitude as
$r_0(\kappa)=\sqrt{R_0(\kappa)}\,e^{i\delta(\kappa)}$. For a real symmetric
single barrier, unitarity gives $|t_0|^2=1-R_0$. Substituting this into the
Fabry--P\'erot transmission amplitude gives
\begin{equation}
|t(\kappa)|^2=
\frac{(1-R_0)^2}
{(1-R_0)^2+4R_0\sin^2[\delta(\kappa)+\kappa d]}.
\label{eq:transmission-resonance-form}
\end{equation}
Consequently, within the Fabry--P\'erot model, unit transmission occurs when
\begin{equation}
\delta(\kappa)+\kappa d=n\pi,\qquad n\in\mathbb{Z}.
\label{eq:resonance}
\end{equation}
The phase $\delta(\kappa)$ is taken as a continuously unwrapped function of
$\kappa$ when solving this condition numerically. Thus, under the symmetry and
unitarity assumptions above, Eq.~(\ref{eq:resonance}) gives $|t|^2=1$ exactly
within the composition model, rather than merely locating a local denominator
minimum. For the full two-centre cusp potential, the resonances are obtained
independently from the direct numerical solution and are slightly displaced
from the Fabry--P\'erot values; Sec.~\ref{sec:validity} quantifies this effect.

\subsection{Connection with the two-channel wire}
\label{sec:multichannel}

The model of Ref.~\cite{Lopez2006} is formulated for a confined quasi-one-dimensional
wire rather than a strictly one-dimensional line. Expanding the wavefunction in
transverse modes, $\psi(x,y)=\sum_n\psi_n(x)\chi_n(y)$, gives
\begin{align}
\left[\frac{d^2}{dx^2}+\kappa_n^2\right]\psi_n(x)
&=\sum_m U_{mn}(x)\psi_m(x),\notag\\
U_{mn}(x)&=\int dy\,\chi_n^*(y)U(x,y)\chi_m(y).
\label{eq:coupledchannels}
\end{align}
For the barrier geometry used here the potential is uniform across the wire,
$U(x,y)=U(x)$. Orthogonality of the transverse functions then gives
$U_{mn}(x)=U(x)\delta_{mn}$, so the open channels decouple. Each longitudinal
component consequently obeys the one-dimensional cusp problem of Sec.~2, with
$\kappa$ replaced by its channel-specific longitudinal wavenumber $\kappa_n$.
The amplitudes $r_0(\kappa_n)$ and $t_0(\kappa_n)$, and hence the double-barrier
amplitudes, therefore apply channel by channel without modification.

This restriction is physically relevant as well as mathematically convenient:
a top gate spanning the wire width is naturally approximated by a transversely
uniform barrier. If the barrier had appreciable transverse structure,
$U_{12}\neq0$ would couple the channels throughout the finite extent of the cusp.
The resulting coupled differential system would no longer reduce to the single
modified-Bessel equation used above and would require a numerical coupled-channel
treatment. We therefore restrict the present analysis to the no-channel-mixing
geometry, which is the regime in which the exact single-cusp amplitudes and the
post-selected two-mode construction can be used directly.

\section{Two-electron concurrence}

\subsection{Setup: two electrons incident on the double barrier}

The analysis below assumes narrow, experimentally resolvable momentum modes
with negligible overlap. Finite-width wave packets would in general require
tracing over unresolved momentum components and would produce a mixed reduced
state; the wave-packet dependence found in Ref.~\cite{Ghanbari2014Wavepackets}
therefore represents a natural extension of the present monochromatic model.

Following~\cite{Lopez2006}, we consider two electrons incident from
the left lead with distinct momenta $\kappa_1,\kappa_2$
($\Delta\kappa=\kappa_2-\kappa_1$), prepared in a common fixed spin state (see Sec.~\ref{sec:spin}), so that
the spin factor is symmetric and the exchange antisymmetry of the two
fermions is carried entirely by the orbital (momentum) degrees of freedom
considered here. Each electron independently scatters off the
double-barrier structure of Sec.~3, coming out either reflected (back
into the left lead) or transmitted (into the right lead) with the
single-particle amplitudes $r(\kappa),t(\kappa)$ derived there. Before
any detection or post-selection, the joint two-particle state is a
product of single-particle scattering states, appropriately
antisymmetrised under exchange of the two fermions.

\subsection{The post-selected two-mode state and its concurrence}

The two-particle final state has three sectors: both electrons
reflected, both transmitted, or one reflected and one transmitted. Following the standard
construction~\cite{Lopez2006}, entanglement between the reflected and
transmitted \emph{modes} is extracted by projecting onto the
one-reflected--one-transmitted sector, in which exactly one electron is
found in the reflected channel and the other in the transmitted
channel. Throughout, ``entanglement'' therefore refers to
mode entanglement between the reflected and transmitted output sectors, not
to entanglement between particle labels, which is not operationally defined
for identical fermions. Labelling the two modes $R$ (reflected) and $T$
(transmitted), the antisymmetry of the two-fermion wavefunction under
exchange of $\kappa_1\leftrightarrow\kappa_2$ together with
$R\leftrightarrow T$ fixes the (unnormalised) post-selected state to
be of the two-mode Bell-like form
\begin{equation}
\ket{\Psi}\;\propto\;t(\kappa_1)\,r(\kappa_2)\,\ket{\kappa_1\!:T,\,
\kappa_2\!:R}\;-\;r(\kappa_1)\,t(\kappa_2)\,\ket{\kappa_1\!:R,\,
\kappa_2\!:T},
\label{eq:postselect}
\end{equation}
i.e.\ a superposition in which the $\kappa_1$-labelled incoming mode is
transmitted while the $\kappa_2$-labelled mode is reflected, and the
converse alternative, with relative amplitude fixed by Fermi statistics. Equation~(\ref{eq:postselect})
is exactly the structure of a two-qubit state
$\alpha\ket{01}-\beta\ket{10}$ with
$\alpha=t(\kappa_1)r(\kappa_2)$, $\beta=r(\kappa_1)t(\kappa_2)$, whose
concurrence~\cite{Wootters1998,Nielsen2000} is the standard result
$\eta=2|\alpha||\beta|/(|\alpha|^2+|\beta|^2)$. Substituting $\alpha,\beta$
gives
\begin{align}
\eta(\kappa_1,\kappa_2)&=\frac{2\,|t(\kappa_1)|\,|r(\kappa_2)|\,
|r(\kappa_1)|\,|t(\kappa_2)|}
{|t(\kappa_1)|^2|r(\kappa_2)|^2+|r(\kappa_1)|^2|t(\kappa_2)|^2}\notag\\
&=\frac{2\,|r(\kappa_1)|\,|t(\kappa_1)|\,
|r(\kappa_2)|\,|t(\kappa_2)|}
{|r(\kappa_1)|^2|t(\kappa_2)|^2+|r(\kappa_2)|^2|t(\kappa_1)|^2},
\label{eq:concurrence}
\end{align}
which depends on the two incident momenta only through the composite
reflection and transmission amplitudes of Sec.~3, evaluated at
$\kappa_1$ and $\kappa_2$ separately, and which is manifestly
symmetric under $\kappa_1\leftrightarrow\kappa_2$ as it must be. A related dependence of generated entanglement on competing reflected and transmitted scattering alternatives appears in analyses of Klein tunnelling in graphene~\cite{Ghanbari2016Graphene}; the microscopic setting differs, but the common structural element is that the scattering amplitudes set the relative weights entering the outgoing entangled state.

The normalisation denominator in Eq.~(\ref{eq:concurrence}) also has a direct
operational meaning.  Before renormalising the one-reflected--one-transmitted
sector, its total probability is
\begin{equation}
P_{RT}(\kappa_1,\kappa_2)
=
|t(\kappa_1)r(\kappa_2)|^2
+
|r(\kappa_1)t(\kappa_2)|^2.
\label{eq:prt}
\end{equation}
Thus $\eta$ quantifies the entanglement conditional on successful projection
onto this sector, whereas $P_{RT}$ quantifies how often that post-selection
event occurs.  A large conditional concurrence therefore need not imply a
large entanglement yield. The conditional concurrence is defined only when
$P_{RT}>0$. If both momenta are simultaneously tuned to perfect-transmission
resonances, then $r(\kappa_1)=r(\kappa_2)=0$ and $P_{RT}=0$; the selected
sector is absent and no normalised conditional state, hence no conditional
concurrence, can be assigned.

\subsection{Relation to entanglement entropy}
\label{sec:entropy}

This short subsection explains why the concurrence used above is also
sufficient to describe the entanglement entropy, a standard quantity used to quantify entanglement in finite
quantum systems. Because $\ket{\Psi}$ of Eq.~(\ref{eq:postselect}) is a
pure state of two modes, its entanglement is completely characterised by
the von Neumann entropy of either single-mode reduced density matrix.
Writing $\alpha=t(\kappa_1)r(\kappa_2)$, $\beta=r(\kappa_1)t(\kappa_2)$
as in Eq.~(\ref{eq:postselect}) and normalising,
the Schmidt coefficients of $\ket{\Psi}$ are
$p=|\alpha|^2/(|\alpha|^2+|\beta|^2)$ and $1-p$, so the reduced density
matrix on either mode has eigenvalues $p,1-p$ and entanglement entropy
\begin{equation}
S=-p\ln p-(1-p)\ln(1-p)\equiv h(p).
\label{eq:entropy}
\end{equation}
Since $\eta=2\sqrt{p(1-p)}$ from Eq.~(\ref{eq:concurrence}), a short
calculation gives $p=\bigl(1+\sqrt{1-\eta^2}\bigr)/2$ (taking, without
loss of generality, the branch $p\ge\tfrac12$, since $h$ is symmetric
under $p\leftrightarrow1-p$), so that the entanglement entropy is a
monotonic function of the concurrence alone,
\begin{equation}
S(\eta)=h\!\left(\frac{1+\sqrt{1-\eta^2}}{2}\right),
\label{eq:entropy_concurrence}
\end{equation}
increasing from $S=0$ at $\eta=0$ (one electron exactly at resonance, see
Sec.~\ref{sec:zeros}) to $S=\ln2$ at $\eta=1$ (a maximally entangled post-selected state). Equation~(\ref{eq:entropy_concurrence}) is the
standard concurrence--entropy relation for a pure two-qubit-like state
(we have verified it numerically against direct diagonalisation of the
reduced density matrix for representative values of $\alpha,\beta$),
and shows that every qualitative statement made below about the
concurrence -- the resonance zeros, the near coincidence of their
locations for the weak barriers considered here, and the changes of the off-resonant concurrence maxima -- carries over to the
entanglement entropy for this pure post-selected state, since $S$ is a strictly increasing function of $\eta$ on $\eta\in[0,1]$. This relation does not apply after tracing over unresolved momentum or spin degrees of freedom, where the reduced state is mixed.

\subsection{Zeros and off-resonant maxima}
\label{sec:zeros}

Within the Fabry--P\'erot model, Eq.~(\ref{eq:concurrence}) vanishes when
exactly one of $\kappa_1$ or $\kappa_2$ satisfies Eq.~(\ref{eq:resonance}),
because one coefficient in Eq.~(\ref{eq:postselect}) then vanishes while
$P_{RT}>0$. If both momenta are simultaneously resonant, $P_{RT}=0$ and the
conditional state is absent rather than separable. For the full two-centre cusp potential the same
mechanism holds at its numerically determined unit-transmission resonances, whose
small displacement from the Fabry--P\'erot values was quantified in
Sec.~\ref{sec:validity}. Exact zeros of $r$ persist there because $U_{\rm 2B}$
is real and even: in the parity basis $t=(e^{2i\delta_+}+e^{2i\delta_-})/2$ and
$r=(e^{2i\delta_+}-e^{2i\delta_-})/2$, with $\delta_\pm$ the even and odd phase
shifts, so $r=0$ requires the single real condition
$\delta_+-\delta_-\in\pi\mathbb{Z}$, which is generically satisfied at
isolated momenta.

Away from resonance, the concurrence reaches unity when the two alternatives in
Eq.~(\ref{eq:postselect}) have equal modulus,
$|t(\kappa_1)r(\kappa_2)|=|r(\kappa_1)t(\kappa_2)|$. Broad maxima therefore
occur between neighbouring transmission resonances. We refer to them below as
\emph{off-resonant concurrence maxima}; the term antiresonance is avoided
because it has a distinct meaning in scattering theory.

\subsection{Role of spin}
\label{sec:spin}

The central results are stated for a common fixed spin state of the two electrons.
Because the barriers are electrostatic and spin independent, an incident creation
operator transforms as
\begin{equation}
c^\dagger_{\kappa,\chi}\longrightarrow
r(\kappa)c^\dagger_{-\kappa,\chi}
+t(\kappa)c^\dagger_{+\kappa,\chi}.
\end{equation}
If both incident electrons have the same spinor $|\chi\rangle$, post-selection on
one reflected and one transmitted particle gives a product of the orbital state
in Eq.~(\ref{eq:postselect}) and the common spin factor. Spin is then a spectator,
and Eq.~(\ref{eq:concurrence}) is exactly the concurrence of the
reflected/transmitted mode state.

For non-identical spinors the situation is more general. The outgoing port
assignment can become correlated with spin; after spin is unobserved or traced out,
the port state is mixed and the pure-state expression Eq.~(\ref{eq:concurrence})
need not apply. The same issue arises for partially overlapping finite momentum
wave packets. A treatment of arbitrary spin overlap or unresolved wave-packet
components therefore requires construction of the reduced density matrix followed
by an appropriate mixed-state entanglement measure. Those extensions are outside
the scope of the present calculation. Restricting the quantitative results to a
common fixed spin state makes the subsystem partition and the meaning of the
reported concurrence unambiguous.

\section{Results}

We first use the principal parameter set $d=2\pi$, $\lambda=0.01$ and
$a=0.3$, for which $d/a\simeq20.94$ and the Fabry--P\'erot approximation has
been independently validated in Sec.~\ref{sec:validity}. The two incident
longitudinal momenta satisfy $\kappa_1>0$, $\kappa_2>0$, and we write
$\Delta\kappa=\kappa_2-\kappa_1>0$ so that the resolved momentum labels are
distinct.

\begin{figure}[htbp]
\centering
\includegraphics[width=0.78\textwidth]{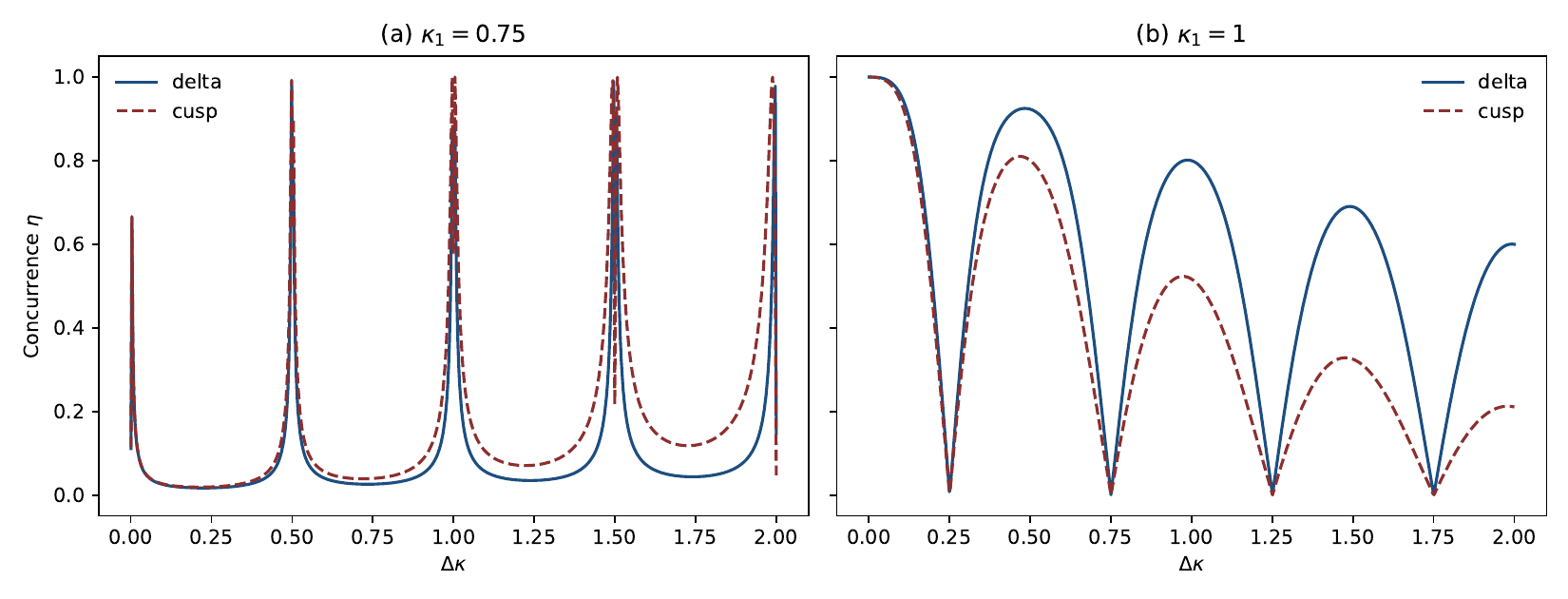}
\caption{Conditional concurrence $\eta(\Delta\kappa)$ in the post-selected
one-reflected--one-transmitted sector for matched-strength double-delta and
cusp barriers, for $a=0.3$, $d=2\pi$, $\lambda=0.01$ and reference momenta
(a) $\kappa_1=0.75$ and (b) $\kappa_1=1$. At these parameters the
Fabry--P\'erot and full two-centre cusp results are indistinguishable on the
scale of the plot (Table~\ref{tab:representative-validation}). The cusp model
retains the same resonance-controlled pattern, while finite width modifies the
locations and, for some reference momenta, the heights of the off-resonant
maxima.}
\label{fig:concurrence}
\end{figure}

Figure~\ref{fig:concurrence} shows the conditional concurrence of the
normalised one-reflected--one-transmitted sector for two representative
reference momenta. The cusp and delta models display the same overall
resonance-controlled structure, but finite barrier width can displace the
off-resonant concurrence maxima and, depending on $\kappa_1$, reduce their
height.

\begin{figure}[htbp]
\centering
\includegraphics[width=0.78\textwidth]{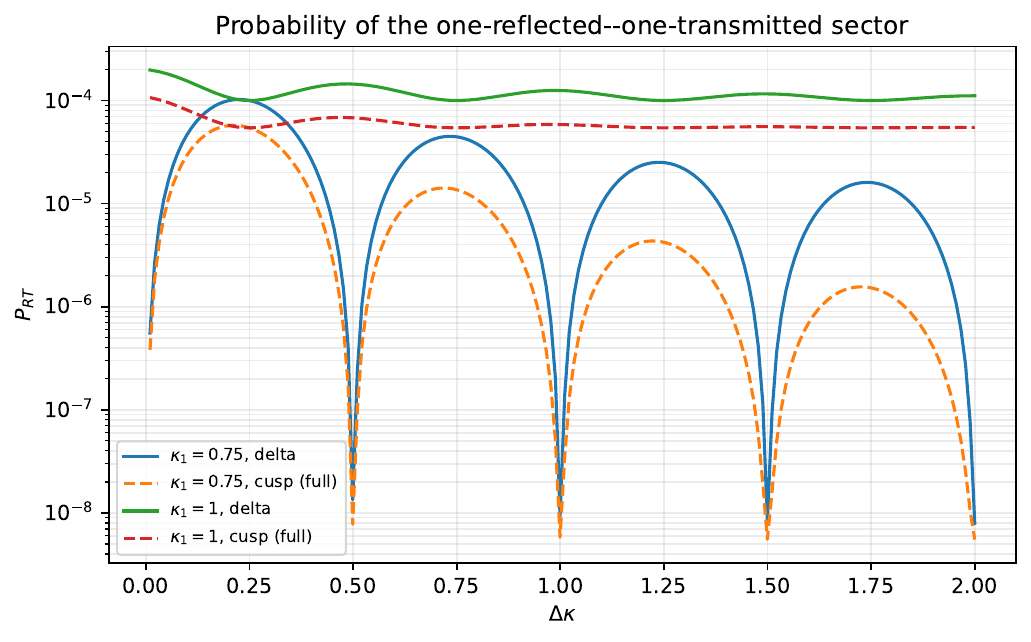}
\caption{Probability $P_{RT}$ of detecting exactly one reflected and one
transmitted electron as a function of $\Delta\kappa$. Solid curves show the
double-delta result and dashed curves the direct numerical solution of the
full two-centre cusp potential for $a=0.3$, $d=2\pi$, $\lambda=0.01$ and
$\kappa_1=0.75$ and $1$.}
\label{fig:prt}
\end{figure}

The conditional nature of this quantity is made explicit by the
post-selection probability $P_{RT}$ defined in Eq.~(\ref{eq:prt}).
Figure~\ref{fig:prt} shows $P_{RT}$ for the same two reference momenta, with
the cusp curves obtained from the full two-centre numerical solution. For
$\kappa_1=1$, the selected-sector probability is of order $10^{-4}$ for the
weak barriers considered here; it can be considerably smaller when the
reference momentum lies close to a transmission resonance. Hence concurrence
close to unity describes strong entanglement conditional on a successful
one-reflected--one-transmitted event and should not be interpreted as
unit-probability production of entangled pairs.

\subsection{Dependence on barrier strength and reference momentum}

To test whether the conclusions are tied to one weak-coupling example, we
repeated the comparison for $\lambda=0.01$, $0.05$ and $0.10$ and for
$\kappa_1=0.75$, $1.00$ and $1.25$, at fixed $a=0.3$ and $d=2\pi$. For each
parameter pair we identify the first three off-resonant concurrence maxima and
define
\begin{equation}
S_j=\frac{\eta^{\rm cusp}_{\max,j}}{\eta^{\delta}_{\max,j}},
\qquad
S_j^{\rm aligned}=
\frac{\eta^{\rm cusp}(\Delta\kappa^{\delta}_{\max,j})}
     {\eta^{\delta}_{\max,j}}.
\label{eq:peak-ratios}
\end{equation}
The first ratio compares corresponding local maxima, while the second
evaluates the cusp concurrence at the momentum difference at which the delta
curve is maximal. Together they separate a displacement of the maximum from
a genuine reduction in its attainable height. The results are listed in Table~\ref{tab:peak-suppression}. The ratios
are essentially independent of $\lambda$ for $\lambda\le0.1$, as expected
from the weak-barrier form-factor result of Sec.~\ref{sec:delta-limit}, in
which the cusp-to-delta reflection ratio $(1+4\kappa^2a^2)^{-1}$ does not
depend on $\lambda$.

\begin{table*}[htbp]
\centering
\caption{Quantitative finite-width modification of the first three
off-resonant concurrence maxima. $S_j$ compares corresponding cusp and delta
maxima, while $S_j^{\rm aligned}$ evaluates the cusp concurrence at the
delta-model peak locations.}
\label{tab:peak-suppression}
\begin{tabular}{cccccc}
\hline
$\lambda$ & $\kappa_1$ & $\langle S_j\rangle$ & $\min S_j$ & $\langle S_j^{\rm aligned}\rangle$ & $\max|\Delta(\Delta\kappa)_{\rm peak}|$\\
\hline
0.01 & 0.75 & 1.0000 & 1.0000 & 0.8483 & $1.06\times10^{-2}$ \\
0.01 & 1.00 & 0.6685 & 0.4763 & 0.6663 & $1.49\times10^{-2}$ \\
0.01 & 1.25 & 1.0000 & 1.0000 & 0.8389 & $4.75\times10^{-3}$ \\
0.05 & 0.75 & 1.0000 & 1.0000 & 0.8454 & $5.28\times10^{-2}$ \\
0.05 & 1.00 & 0.6671 & 0.4749 & 0.6649 & $1.49\times10^{-2}$ \\
0.05 & 1.25 & 1.0000 & 1.0000 & 0.8363 & $1.63\times10^{-2}$ \\
0.10 & 0.75 & 1.0000 & 1.0000 & 0.8360 & $5.57\times10^{-2}$ \\
0.10 & 1.00 & 0.6654 & 0.4732 & 0.6633 & $1.48\times10^{-2}$ \\
0.10 & 1.25 & 1.0000 & 1.0000 & 0.8315 & $1.93\times10^{-2}$ \\
\hline
\end{tabular}
\end{table*}

The scan shows that finite width is not represented by a universal
suppression factor. For $\kappa_1=1$ the corresponding cusp maxima are
substantially lower than the delta maxima, whereas for the two reference
momenta close to resonance both models can still attain concurrence near
unity and the dominant finite-width effect is a displacement of the maxima.
Finite width therefore does not simply suppress every off-resonant peak;
whether it displaces or lowers the maxima depends on the reference momentum.

The physical dependence on the cusp width itself was also quantified at
fixed $\lambda=0.01$, $\kappa_1=1$ and $d=2\pi$. For each $a$ we evaluated
the first three corresponding off-resonant maxima, using the full two-centre
solution at the cusp peak positions. Table~\ref{tab:a-dependence} shows that
the mean peak ratio decreases systematically as the barrier is broadened.
Thus the reduction of the maxima seen in Fig.~\ref{fig:concurrence}(b)
grows systematically with the barrier width: increasing the spatial extent of
the cusp lowers the attainable conditional concurrence for this non-resonant
reference momentum.

\begin{table}[htbp]
\centering
\caption{Dependence of the first three off-resonant concurrence maxima on
cusp width for $\lambda=0.01$, $\kappa_1=1$ and $d=2\pi$. The table reports
$\langle S_j\rangle=\langle\eta^{\rm cusp}_{\max,j}/\eta^{\delta}_{\max,j}\rangle$
from the full two-centre cusp calculation.}
\label{tab:a-dependence}
\begin{tabular}{ccc}
\hline
$a$ & $d/a$ & $\langle S_j\rangle$ \\
\hline
0.1 & 62.83 & 0.929 \\
0.2 & 31.42 & 0.789 \\
0.3 & 20.94 & 0.668 \\
0.6 & 10.47 & 0.490 \\
0.9 & 6.98  & 0.431 \\
\hline
\end{tabular}
\end{table}

The first three full-cusp resonance positions were also computed as
$\lambda$ was varied. The largest cusp--delta displacement among these
resonances grows with barrier strength, while the difference between the full
two-centre cusp calculation and the Fabry--P\'erot approximation remains much
smaller in the well-separated geometry considered here.

\begin{table}[htbp]
\centering
\caption{Largest cusp--delta resonance shift among the first three
transmission resonances for $a=0.3$ and $d=2\pi$.}
\label{tab:lambda-resonance-shifts}
\begin{tabular}{cc}
\hline
$\lambda$ & $\max_{n=1,2,3}|\kappa_{n,\rm cusp}-\kappa_{n,\delta}|$\\
\hline
0.01 & $6.75\times10^{-6}$ \\ 
0.05 & $1.56\times10^{-4}$ \\ 
0.10 & $5.67\times10^{-4}$ \\ 
\hline
\end{tabular}
\end{table}

\section{Conclusion}

We have examined whether the resonance-controlled conditional entanglement found
for ideal double-delta barriers survives when each scatterer has a finite spatial
width. For the exponentially localised cusp barrier the single-scatterer problem
is exactly solvable in terms of modified Bessel functions of imaginary order. We
derived the asymptotic normalisation and the derivative matching at the cusp,
obtained closed-form amplitudes, verified current conservation analytically, and
checked the complex amplitudes independently by direct numerical integration of
the Schr\"odinger equation.

At fixed integrated strength $\lambda=2U_0a$, the cusp family approaches
$\lambda\delta(x)$ distributionally. Expanding the exact amplitudes shows that
the first finite-width correction to both $r_0$ and $t_0$ is $O(a)$, not $O(a^2)$;
the numerical log--log study confirms this asymptotic law. For the double cusp,
the usual Fabry--P\'erot composition is not exact because the exponential tails
overlap. Direct integration of the full two-centre potential shows, however, that
for the principal geometry $d/a\simeq20.94$ the first three resonance positions
agree with the composition model to within $1.6\times10^{-9}$ in $\kappa$, with the error
increasing as $d/a$ is reduced. The well-separated-scatterer approximation is
therefore quantitatively controlled in the regime used for the entanglement plots.

Within that regime, the resonance condition
$\delta(\kappa)+\kappa d=n\pi$ produces unit transmission in the composition
model, while the full two-centre resonances are displaced only slightly. When exactly one incident momentum is at a unit-transmission resonance,
the one-reflected--one-transmitted post-selected state contains only one
alternative, so its conditional concurrence vanishes. If both momenta are
simultaneously resonant, the selected-sector probability is zero and the
conditional concurrence is undefined. Between
resonances, finite barrier width can both displace the concurrence maxima and,
depending on the reference momentum, reduce their attainable height. The scan in
$\lambda$ and $\kappa_1$ shows that these two effects should not be conflated into
a universal suppression factor.

The post-selection probability $P_{RT}$ is an essential part of this interpretation.
For the weak barriers used in the principal examples it is typically small, so
concurrence close to unity means strong entanglement \emph{conditional on} observing
one reflected and one transmitted electron, not deterministic production of an
entangled pair. The quantitative conclusions are correspondingly restricted to
resolved, narrow momentum modes, a common fixed spin state, and the weak-barrier,
well-separated parameter regime tested here. Finite-width wave packets, arbitrary
spin overlap, channel mixing and finite temperature generally produce mixed reduced
states and require a density-matrix treatment. In particular, at finite temperature
one must first form the thermally averaged outgoing density matrix and then evaluate
a suitable mixed-state entanglement measure; averaging the zero-temperature pure-state
concurrence or entropy is not equivalent. These extensions provide natural directions
for future work, together with inelastic scattering and electron--electron interactions.

\section*{Conflict of interest}

The author declares that he has no known competing financial interests or
personal relationships that could have appeared to influence the work
reported in this paper.

\section*{Ethics statement}

This work did not involve human participants, animals, or clinical data.
No ethics approval was required.

\section*{Funding}
The author received no specific funding for this work.

\section*{Data availability}
No external data were used in this work. The numerical data shown in the figures and tables were generated from
the analytical expressions and direct numerical integrations described in
the article and are available from the author upon reasonable request.


\begin{thebibliography}{99}

\bibitem{Martin2003} Martin T, Cr\'epieux A and Chtchelkatchev N 2003
in \textit{Quantum Noise in Mesoscopic Physics}, ed Nazarov Y V
(Dordrecht: Kluwer) NATO Science Series II 97

\bibitem{Beenakker2003} Beenakker C W J, Emary C, Kindermann M and
van Velsen J L 2003 Phys. Rev. Lett. 91 147901

\bibitem{Schliemann2001} Schliemann J, Loss D and MacDonald A H 2001
Phys. Rev. B 63 085311

\bibitem{Edlbauer2022} Edlbauer H \textit{et al.} 2022 EPJ Quantum
Technol. 9 21

\bibitem{Ubbelohde2023} Ubbelohde N, Freise L, Pavlovska E, Silvestrov
P G, Recher P, Kokainis M, Barinovs G, Hohls F, Weimann T, Pierz K and
Kashcheyevs V 2023 Nat. Nanotechnol. 18 733

\bibitem{Ryu2022} Ryu S and Sim H-S 2022 Phys. Rev. Lett. 129 166801

\bibitem{Ghanbari2013Spin} Ghanbari-Adivi E, Soltani M and Ebtekarnasab H 2013
Eur. Phys. J. D 67 118

\bibitem{Ghanbari2014Wavepackets} Ghanbari-Adivi E, Soltani M and
Ebtekarnasab H 2014 Eur. Phys. J. D 68 103

\bibitem{Ghanbari2014Collisions} Ghanbari-Adivi E and Soltani M 2014
Eur. Phys. J. D 68 336

\bibitem{Ghanbari2016Waveguides} Ghanbari-Adivi E, Soltani M, Alami Z and
Sheikhali M 2016 Quantum Inf. Process. 15 4219--4236

\bibitem{Ghanbari2016Graphene} Ghanbari-Adivi E, Soltani M and Sheikhali M
2016 Quantum Inf. Process. 15 2377--2391

\bibitem{Lopez2006} L\'opez A, Rend\'on O, Villalba V M and Medina E
2007 Phys. Rev. B 75 033401

\bibitem{Razavy2003} Razavy M 2003 \textit{Quantum Theory of Tunneling}
(Singapore: World Scientific)

\bibitem{MelloKumar2004} Mello P A and Kumar N 2004 \textit{Quantum
Transport in Mesoscopic Systems: Complexity and Statistical Fluctuations}
(Oxford: Oxford University Press)

\bibitem{Wootters1998} Wootters W K 1998 Phys. Rev. Lett. 80 2245

\bibitem{Nielsen2000} Nielsen M A and Chuang I L 2000 \textit{Quantum
Computation and Quantum Information} (Cambridge: Cambridge University
Press)

\end{thebibliography}
\end{document}